\documentclass[final,supergroupaddress]{siamltex}

\usepackage{graphicx}% Include figure files
\usepackage{dcolumn}% Align table columns on decimal point
\usepackage{bm}% bold math
\usepackage{color}
\usepackage{epsfig}
\usepackage{multirow}
\usepackage{amsmath,amssymb}

\title{Mellin Transform and Image Charge Method for Dielectric Sphere in an Electrolyte}

\author{Zhenli Xu\thanks{Department of Mathematics, Institute of Natural Sciences, and MoE Key Lab for Scientific and Engineering Computing, Shanghai Jiao Tong University, Shanghai 200240, China. Email: xuzl@sjtu.edu.cn}
        \and Yihao Liang\thanks{Department of Physics, and Institute of Natural Sciences, Shanghai Jiao Tong University, Shanghai 200240, China. Email: liangyihao@sjtu.edu.cn}
        \and Xiangjun Xing \thanks{Department of Physics, and Institute of Natural Sciences, Shanghai Jiao Tong University, Shanghai 200240, China. Email: xxing@sjtu.edu.cn}}
\begin{document}

\maketitle

\begin{abstract}
We revisit the image charge method for the Green's function problem of the Poisson-Boltzmann equation for a dielectric sphere immersed in ionic solutions.  Using finite Mellin transformation, we represent the reaction potential due to a source charge inside the sphere in terms of one dimensional distribution of image charges.  The image charges are generically composed of a point image at the Kelvin point and a line image extending from the Kelvin point to infinity with an oscillatory line charge strength.   We further develop an efficient and accurate algorithm for discretization of the line image using Pad\'e approximation and finite fraction expansion. Finally we illustrate the power of our method by applying it in a multiscale reaction-field Monte Carlo simulation of monovalent electrolytes.

% This new expression with the image representation can be more rapidly calculated in comparison to the Kirkwood series with the spherical harmonics, and is useful in computer simulations with multiscale reaction field models in different applications such as the hybrid explicit/implicit model for biomolecular solvation and the hybrid primitive/implicit model for colloidal suspensions. Numerical demonstrations are present to show the necessarity and attractive feature of the image-based reaction field in simulations of ionic fluids, where two examples are used: one is to calculate self energies of point charges compared to the spherical harmonic expansion, and the other one is to use Monte Carlo simulations for electrolytes.
\end{abstract}



\pagestyle{myheadings}
\thispagestyle{plain}
\markboth{Z. Xu, Y. Liang and X. Xing}{Image Charge Method for Electrolyte}

\section{Introduction}

The method of image charges is a classical technique \cite{Smythe:book:89,Jackson:book:01} for electrostatic problems.  Its most elementary application is the problem of a point charge in a spherical cavity inside a conducting medium (or the reciprocal problem of a point charge outside a conducting sphere).  In 1845, William Thomson (Lord Kelvin) \cite{Thomson:JMPA:1845}  noticed that the vanishing-potential boundary condition of conductors can be automatically satisfied on the sphere by putting an image point charge at the Kelvin point.
A natural extension is the problem of a dielectric sphere inside a different dielectric background, where a single point image charge no longer works.
In 1883 Carl Neumann \cite{Neumann:TL:83} discovered that a point image at the Kelvin point together with a line image starting from the Kelvin point to infinity solves the boundary condition. This result has been independently re-derived by several authors, up to 1990s; see the reviews \cite{OL:RSB:03,XC:SIREV:11} for more details.
More recently, the problem of a spherical cavity inside an ionic solution has been studied \cite{DC:CCP:07,XDC:JCP:09}.  It is again found that the image charge distribution consists of a point image and a line image.  Computational method for the line charge density has been developed which works well in the asymptotic limits \cite{DC:CCP:07,XDC:JCP:09} that $\kappa R$ is either large or small, where $\kappa$ is the inverse Debye length, while $R$ is the radius of spherical cavity.   The current work addresses the general case where $\kappa R$ is neither large nor small. % and a general dielectric ratio between inside and outside.

The main advantage of image charge methods is to represent the effects of polarization charges in terms of point charges or line of charges, and therefore avoiding the task of numerically solving boundary value problems.  This can substantially reduce the computational cost in Monte Carlo or molecular dynamics simulation of charged systems.
Further reduction of computational cost can be achieved by discretizing line images using Gaussian quadratures, which effectively approximates a  line charge by a few point charges. The total electrostatic energy of the system can then be represented as Coulomb interaction between many point charges.

\begin{figure*}[htp!]
\centering %\includegraphics[scale=.35]{model1.eps}
\includegraphics[scale=.45]{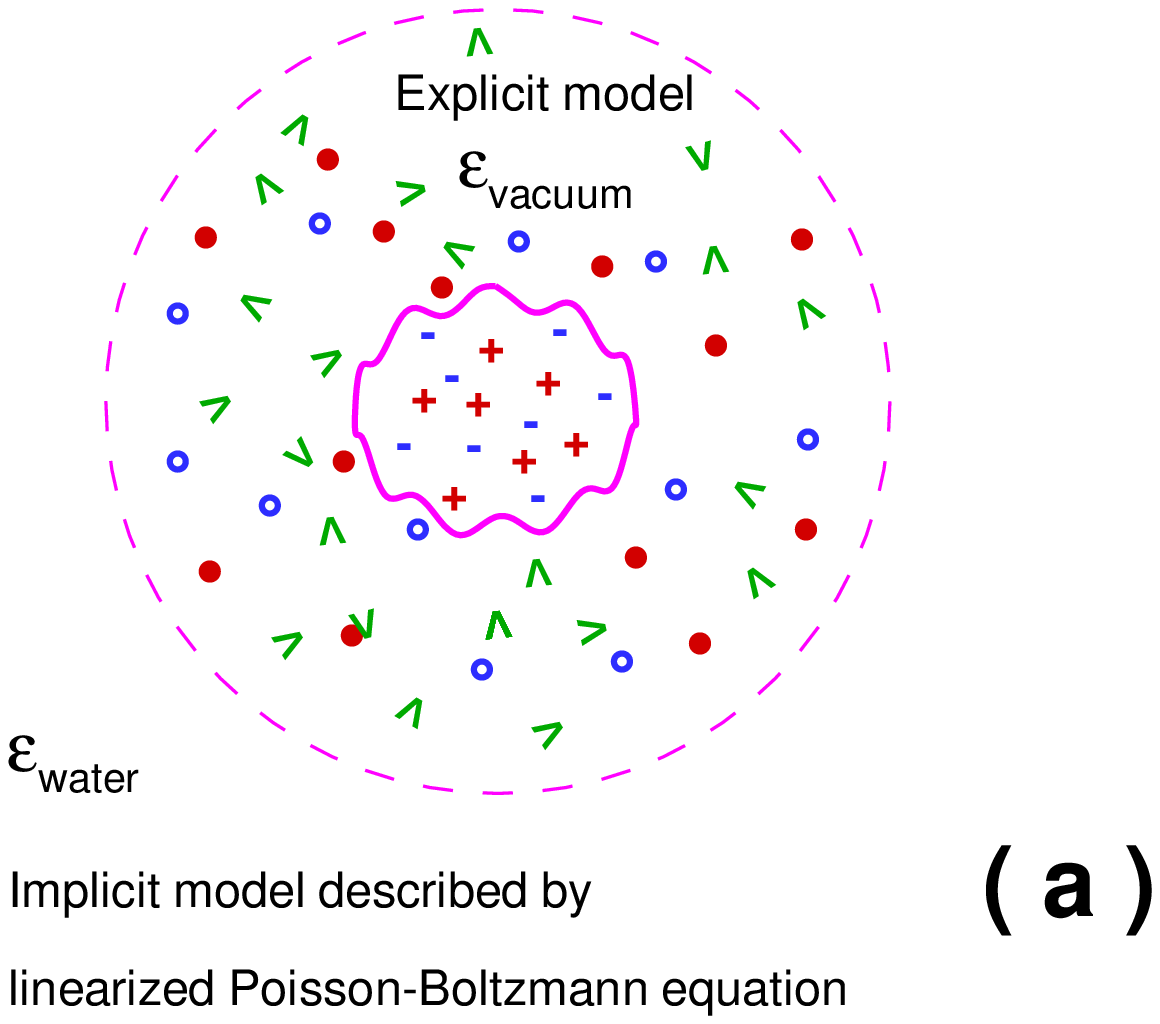}\includegraphics[scale=.45]{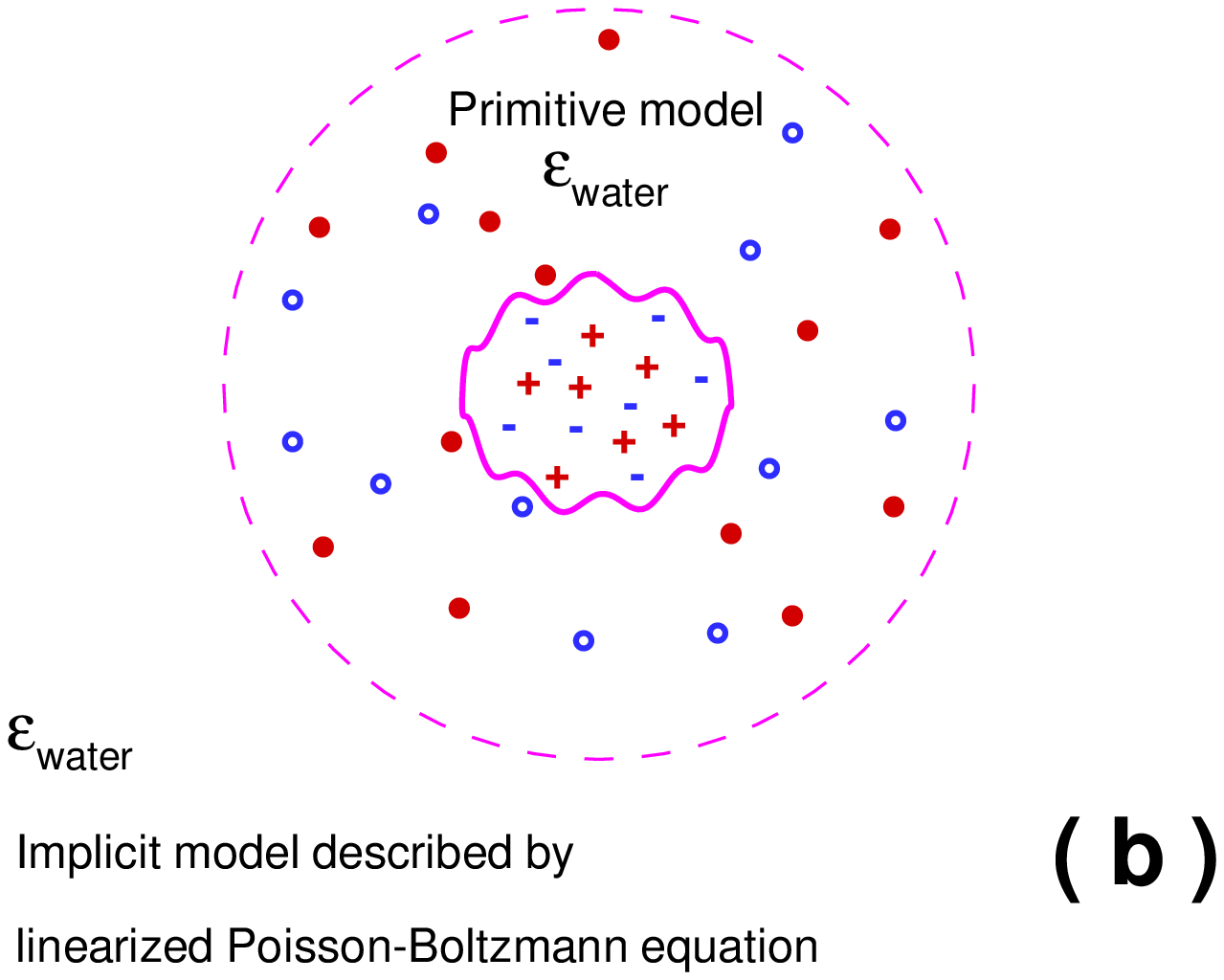}
\caption{Schematic illustrations of two multi-scale models of solute-solvent systems. The plus and minus signs represent the charges in the solute,  the circles represent mobile ions in the solvent, and the angles represent the explicit water molecules. (a) The hybrid explicit/implicit solvent model.  Inside the cavity, both water molecules and ions are treated explicitly. Outside the cavity, both bulk solvent and the ions are treated using the continuum theory; (b) The hybrid primitive/implicit solvent model.  Inside the cavity, the water molecules are treated implicitly, while the ions are treated explicitly, i.e. the so-called primitive model. Outside the cavity, everything is treated implicitly. } \label{schem}
\end{figure*}

Most interfaces appearing in nature have irregular shapes, and the corresponding Green's functions can not be represented in terms of simple distributions of image charges.  Hence one may rightfully argue that image charge methods have rather limited applications.   The most important usage of image charge methods, however, arises in the multiscale reaction-field modeling of Coulomb many body systems, where the interfaces are {\em artificially introduced}.  Spherical interfaces are almost always used for their simplicity, and for the availability of analytic results for image charges.  Research along this direction is of current interest in both simulations and continuum modeling \cite{OS:ARCC:06,EHL:JCP:10,Messina:JPCM:09,FPPR:RMP:10,WKDG:NS:11,BSK:PRL:11,WZCX:SIREV:12}.
 
Due to the long range nature of Coulomb interaction, simulation of charged systems is highly nontrivial.  A proper treatment of boundary conditions is vital in order to obtain physically meaningful results. Periodic boundary conditions can remove artificial boundary effects in a self-consistent fashion and restore the translation symmetry.  Combined with Ewald summation method \cite{Ewald:AP:21}, the cost of computing the total energy of the system with $N$ particles scale as $N^{3/2}$.  This can be further reduced to the order of $N\log N$ using a mesh-based algorithm such as the particle mesh Ewald or particle-particle particle-mesh Ewald lattice summation techniques.  The periodic images are however unphysical and may produce artifacts that obscure the real physics.  Besides, computational cost of Ewald-type summation method is still rather prohibitive for  large systems in Monte Carlo simulations, which limits simulation of charged systems to rather small size. The development of non-Eward methods remains important topics; See Fukuda and Nakamura \cite{FN:BR:12} for a recent review.

An attractive alternative is to use the reaction field type of modeling, which is essentially a multi-scale strategy, schematically illustrated in Figure \ref{schem}.  In this approach, an artificial (spherical) cavity is introduced.  All ions inside the cavity, together with possible mesoscopic objects such as charged proteins and colloids, are treated explicitly {using microscopic model (such as the primitive model \cite{Linse:APS:05}) together with Monte Carlo/Molecular Dynamics}\,\footnote{Strictly speaking, grand canonical ensemble must be used in order to treat charge fluctuations properly. }, while ions outside the cavity are treated implicitly using appropriate continuum theory, such as linearized Poisson-Boltzmann (PB) theory.  It is known that linearized PB provides an accurate approximation to dilute electrolytes.  For any charge inside the cavity, then, we must solve the electrostatic Green's function problem, where the potential satisfies Poisson equation inside the cavity, and satisfies the (linearized) PB equation outside. {The resulted Green's function provides the pairwise electrostatic interaction (the force field) between mobile ions}.  This problem can be efficiently solved using the image charge method developed in this work.

In this multi-scale modeling approach, the microscopic model inside the system and the continuous theory outside the cavity really describe the same system.  Therefore the parameters of the Poisson-Boltzmann theory need to be determined self-consistently.  These include the Debye length and the effective dielectric constant.  For dilute electrolyte, the Debye length can be theoretically calculated as a function of ion densities, while the dielectric constant can be taken to be that of the solvent.  Some unphysical artifacts also arise because of the artificial hard wall repulsion of the cavity surface.  Using statistical mechanics, one can study this artifact and use extra short range interaction to compensate it.  Alternatively, one can also ignore the details inside a thin shell near the cavity surface of thickness 1 - 2 ion diameters.  Finally, to achieve the balance between  efficiency and precision, the radius of cavity should be chosen to be couple of the Debye length.  We also note that there can be different levels of modeling inside the cavity.  In the so-called hybrid implicit/explicit model \cite{OS:ARCC:06}, both solvent molecules and ions are treated explicitly inside the cavity.  By contrast, in the so-called hybrid primitive/implicit model, the solvent inside the cavity is modeled implicitly as a dielectric medium, while the ions are treated at the level of primitive model.

The reaction field model augmented by our image charge methods is therefore able to provide an accurate treatment of the electrostatic boundary conditions.  Its computation cost of the total energy scales as $N^2$, where $N$ here is the total number of particles including all ions as well as their images.  While for small systems (with total particle number typically smaller than 1000) this cost is manageable, for large systems, it becomes unrealistic.   Fortunately, the computation cost can be dramatically reduced using fast multipole based methods \cite{GR:JCP:87,BH:Nature:86,DK:JCP:00,GR:AN:97}.  The computational cost for the combined method generally scales as $N \log N$.

In this paper, we mainly focus on the image charge method for the Green's function problem of these multi-scale reaction field models.  The statistical mechanical foundation of these reaction field models will be discussed in a separate publication.  In the remaining of this work, we shall first define the Kirkwood series for the Green's function (Sec. II) and then use the inverse Mellin transform to find the image charge representation (Sec. III).  We further discretize the line image using method of Gauss quadrature and construct an efficient numerical scheme for the computation of the reaction potential.  In Sec. IV, we compute the reaction potential using our method and quantify the errors.  We also demonstrate the power of multi-scale modeling by Monte Carlo simulating a dilute symmetric electrolyte.

\section{Green's function of the Poisson-Boltzmann equation}

\begin{figure}[t!]
\centering\includegraphics[scale=.4]{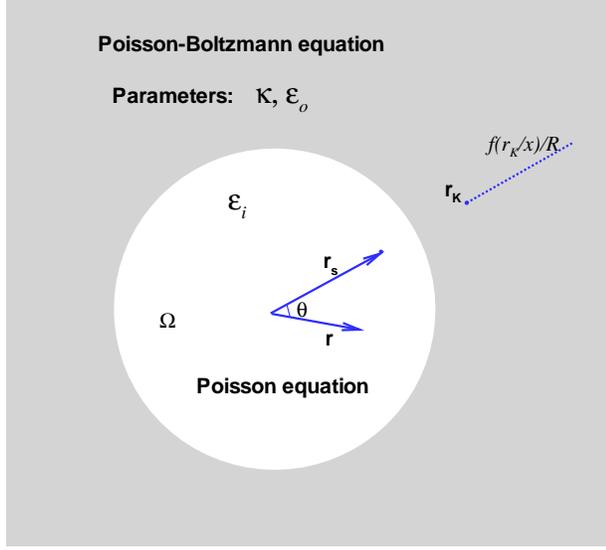}
\caption{The geometry of the Green's function problem for a point charge at the source point $\mathbf{r}_s$.  The reaction potential at a field point $\mathbf{r}$ is generated by a point image at the Kelvin point $\mathbf{r}_K$ and a line image extends from $\mathbf{r}_K$ to infinity.  The line charge density is given by Theorem \ref{theom}. } \label{schem2}
\end{figure}

As illustrated in Figure \ref{schem2}, let $\Omega\in \mathbb{R}^3$  be a spherical cavity with radius $R$, centered at the origin,  with dielectric constant $\varepsilon_\mathrm{i}$.  The volume outside the cavity is filled with electrolyte, which is described by linearized Poisson-Boltzmann theory and is characterized by a Debye length $\kappa^{-1}$ and dielectric constant $\varepsilon_\mathrm{o}$.  Consider a point unit charge fixed at $\mathbf{r}_s$ inside the cavity, the average potential (averaged over the statistical fluctuations of electrolyte outside the cavity) satisfies the following equations:
%the relevant electrostatic Green's function satisfies the following equations:
\begin{equation}\label{green}
\left\{\begin{array}{ll}
-\varepsilon_\mathrm{i} \nabla^2G_\mathrm{i}(\mathbf{r},\mathbf{r}_s)
=\delta(\mathbf{r}-\mathbf{r}_s), \quad \quad\quad~~\mathbf{r}\in\Omega,
\vspace{3mm}\\
- \nabla^2G_\mathrm{o}(\mathbf{r},\mathbf{r}_s)
+ \kappa^2G_\mathrm{o}(\mathbf{r},\mathbf{r}_s)= 0
,~~~~\mathbf{r}\not\in\Omega,
\end{array}\right.
\end{equation}
where $G_\mathrm{i}$ and $G_\mathrm{o}$ are the average potential inside and outside the cavity respectively.  The boundary condition at infinity is
\begin{equation}
G_\mathrm{o}(\mathbf{r},\mathbf{r}_s)\rightarrow 0, ~~~\hbox{for} ~~ r\rightarrow\infty.
\end{equation}
Throughout the paper we use light italic letter $p$ to represent the magnitude of a vector $\mathbf{p}$. On the cavity boundary, the Green's function satisfies the following standard electrostatic interface conditions:
 \begin{equation}
G_\mathrm{i}=G_\mathrm{o}, \quad
\varepsilon_\mathrm{i}\partial_rG_\mathrm{i}
=\varepsilon_\mathrm{o}\partial_rG_\mathrm{o}, ~~~\hbox{at} ~~ r=R,
\label{boundary conditions}
\end{equation}
The general boundary value problem associated with Eqs. \eqref{green}-\eqref{boundary conditions} actually defines the electrostatic Green's function,
$G(\mathbf{r},\mathbf{r}_s)$, which equals $G_\mathrm{i}$ inside the cavity and $G_\mathrm{o}$ outside. It is this Green's function that shall be directly used in the multi-scale reaction field modeling of electrolyte.  The inverse Debye length is defined by $\kappa=(4\pi\l_B\sum_j \rho_j^0z_j^2)^{1/2},$
where $l_B$ is the Bjerrum length of the solvent ($l_B=7.14 \AA$ for water at room temperature), $\rho_j^0$ and $z_j$ are the bulk concentration and the valence of the $j$th species of ions.

The Green's function defined in Eqs.~\eqref{green}-\eqref{boundary conditions} has an azimuthal symmetry, hence depends only on $r$ and $\theta$, where $\theta$ is the angle between the source point $\mathbf{r}_s$ and the field point $\mathbf{r}$, see Fig.~\ref{schem2}.
The potential inside the cavity can be written as,
\begin{equation}
G_\mathrm{i}=\Phi_\mathrm{coul}+\Phi_\mathrm{rf},
\end{equation}
which is the superposition of the direct Coulomb potential $\Phi_\mathrm{coul}=1/(4\pi\varepsilon_\mathrm{i}|\mathbf{r}-\mathbf{r}_s|)$ and the reaction potential $\Phi_\mathrm{rf}$ which is a harmonic function.  Both potentials can be expanded in terms of spherical harmonics:
\begin{subequations}
\label{potentials-all}
\begin{eqnarray}
\frac{1}{|\mathbf{r}-\mathbf{r}_s|} &=&
%\frac{1}{4\pi\varepsilon_\mathrm{i}}
\sum_{n=0}^\infty \frac{r_<^n}{r_>^{n+1}}P_n(\cos\theta),
\label{potentials-all-1}\\
\Phi_\mathrm{rf}(\mathbf{r}, \mathbf{r}_s) &=&
\sum_{n=0}^\infty A_n r^n P_n(\cos\theta),
\end{eqnarray}
where $r_< (r_>)$ is the smaller (larger) one between $r_s$ and $r$,  while $A_n$ are constants to be determined,
and $P_n(\cdot)$ is Legendre polynomial of order $n$.
The potential outside the cavity can also be expanded in terms of Legendre polynomials:
\begin{eqnarray}
\Phi_\mathrm{o} &=&
\sum_{n=0}^\infty B_n k_n(\kappa r) P_n(\cos\theta),
\end{eqnarray}
\end{subequations}
where $k_n(\cdot)$ is the {\it modified spherical Hankel function} (also called the modified spherical Bessel function of the third kind)  \cite{AS:book:64}, defined by the following series,
\begin{equation}
k_{n}(u)=\frac{\pi \mathrm{e}^{-u}}{2u}
\sum_{l=0}^{n}\frac{(n+l)!}{l!(n-l)!}
\frac {1}{(2u)^{l}}.
\label{knu-def}
\end{equation}

The coefficients $A_n, B_n$ in Eqs.~(\ref{potentials-all}) can be found by solving the standard electrostatic boundary conditions Eq.~(\ref{boundary conditions}).   We shall define the Kelvin point $\mathbf{r}_K$ via
\begin{equation}
\mathbf{r}_K=(R/r_s)^2\mathbf{r}_s.
\label{Kelvin-point-def}
\end{equation}
For source charge inside the cavity, we always have
\begin{equation}
r_s < R < {R^2}/{r_s} = r_K.
\end{equation}
Using the orthogonality of the spherical harmonics, we obtain the following expansion of the reaction potential, widely known as {\em Kirkwood series expansion} \cite{Kirkwood:JCP:34}:
\begin{equation} \label{kirkwood}
\Phi_\mathrm{rf} =
\frac{R/r_s}{4\pi\varepsilon_\mathrm{i}}
\sum_{n=0}^\infty \frac{r^n}{r_K^{n+1}}
M_n(u) P_n(\cos\theta),
\end{equation}
with the harmonic coefficients given by
\begin{equation}
M_n(u)=\frac{\varepsilon (n+1)k_n(u)+uk_n'(u)}
{\varepsilon nk_n(u)-uk_n'(u)}, \label{kernel}
\end{equation}
where $\varepsilon=\varepsilon_\mathrm{i}/\varepsilon_\mathrm{o}$ and $u=\kappa R.$ The
Kirkwood series expansion has been used in the calculation of the reaction potential inside a spherical cavity in simulations \cite{KW:JCP:89,BR:JCP:94}.  It converges slowly when a source charge approaches to the cavity surface, which prevents its wide application in dynamical and Monte Carlo simulations.

\section{Image charge representation and algorithm}
In this section, we develop an image charge representation for the reaction potential using finite Mellin transformation.

\subsection{Finite Mellin transform}

Let $f(t)$ be a function defined in the interval $[0,1]$. Its {\em finite Mellin transform}, $F(s)$ is defined as \cite{GR:book:07}
\begin{equation}
F(s)=\mathcal{M}[f(t);s] = \int_0^1 f(t) \, t^{s-1}dt,
\end{equation}
where $s$ is generically a complex variable.
The original function $f(t)$ can be expressed in terms of $F(s)$ by the {\em inverse Mellin transform}
\begin{equation}
f(t)=\mathcal{M}^{-1}[F(s);t]
=\int_{\sigma-i\infty}^{\sigma+i\infty} F(s)\, t^{-s}ds.
\end{equation}
Here the integration is carried out over a vertical line, $Re(s)=\sigma$, in  the complex plane. The dual functions $\{ f(t), F(s)\}$ form a {\em finite Mellin transform pair}.
%The finite Mellin transform and its inversion exist when the integral $\int_0^1 |f(t)|t^{\alpha-1}dt$ is bounded for some $\alpha>0$ and $\sigma>\alpha.$

The finite Mellin transform is closely related to the one-sided Laplace transform.
Let $t=e^{-x}$, then $\mathcal{M}[f(t); s]=\mathcal{L}[f(e^{-x});s]. $
Likewise, the inverse transform can also be expressed in terms of the inverse Laplace transform.  The finite Mellin transform is equivalent to the usual Mellin transform for the function with compact support in the finite interval. The usual Mellin transform can be defined in terms of the two-sided Laplace transform.

Two finite Mellin transform pairs shall be useful in our discussion below,
\begin{subequations}
\label{pair}
\begin{eqnarray}
 \{ \delta(t-1),\quad1\},
 \quad
 \left\{ t^c,\quad {(s+c)^{-1}}  \right\},
\end{eqnarray}
\end{subequations}
where $\delta$ is the Dirac delta function.  The Mellin transformation of the second pair is defined for $Re(s+c)>0$.

We shall consider $M_n(u)$ in  Eq.~(\ref{kernel}) as a function of $n$ and analytically continue the integer variable $n$ into the complex plane.  This can be done through using the integral representation of Bessel functions, for which $k_n(u)=\sqrt{\pi/2u}K_{n+1/2}(u)$ and the modified Bessel function of the second kind (see, e.g. Ref. \cite{GR:book:07}, pp. 917)
\begin{equation}K_\nu(u)=\frac{\pi}{2}\left(\frac{u}{2}\right)^\nu\int_0^\infty\frac{e^{-z-u^2/4z}dz}{z^{\nu+1}},\end{equation}
for $u>0$.  Let $f(t)$ be the inverse finite Mellin transformation of $M_n(u)$,\footnote{ We shall suppress the dependence of $f(t)$ on $u$, to avoid cluttered notations.} we have
\begin{subequations}
\begin{eqnarray}
f(t) &=& \mathcal{M}^{-1}(M_n(u); t),\\
M_n(u) &=&\mathcal{M}(f; n)
= \int_0^1 f(t) \, t^{n-1}dt.
\label{Mellin-2}
\end{eqnarray}
\end{subequations}
As given by Eq.~(\ref{kernel}), the harmonic coefficient $M_n(u)$ is finite for all $Re(n)\geq0$, which ensures the existence of the finite Mellin transform.  Substituting Eq.~(\ref{Mellin-2}) into the Kirkwood series Eq.~(\ref{kirkwood}), and using $r_K r_s = R^2$, the reaction potential can be re-expressed as
\begin{equation}
\Phi_\mathrm{rf} =
\frac{1}{4\pi R\varepsilon_\mathrm{i}}\int_0^1 dt
\frac{r_K f(t)}{t^2} \sum_{n=0}^\infty
\frac{r^n }{\left(r_K/t\right)^{n+1}}
P_n(\cos\theta) . \label{rf}
\end{equation}
Let us further define a vector $\mathbf{x} = \mathbf{r}_K/t$.  As $t$ decreases from $1$ to $0$, the vector $\mathbf{x}$ runs from the Kelvin point $r_K$ to infinity along the radial direction.  We shall see that this is precisely the loci of the image charge line.  Let $x$ be the magnitude of vector $\mathbf{x}$, we have $x=r_K/t$.  Since we are only interested in the field point inside the cavity, we have $r \leq R \leq r_K/t = x $, hence we can sum the series in Eq.~(\ref{rf}) using the expansion Eq.~(\ref{potentials-all-1}) and obtain
\begin{eqnarray}
\label{image1}
\Phi_\mathrm{rf}
%&=& \frac{1}{4\pi R\varepsilon_\mathrm{i}}
%\int_0^1 dt \frac{r_K f(t)}
% {t^2|\mathbf{r}-\mathbf{r}_K/t|}\nonumber\\
= \frac{1}{4\pi \varepsilon_\mathrm{i}}
\int_{r_K}^\infty dx \frac{R^{-1} \,f(r_K/x)}{|\mathbf{r}-\mathbf{x}|}.
\end{eqnarray}
The last integral represents the potential generated by one dimensional distribution of image charges along the radial direction, which starts from the Kelvin point $\mathbf{r}_K$ and extends to infinity.  The linear charge density is $\rho(x) = R^{-1} f(r_K/x) $.  The geometry is illustrated in Fig.~\ref{schem2}.   Therefore we arrive at our main result in this work.

\begin{theorem} \label{theom}
Let $\{f(t),M_n(u)\}$ be a finite Mellin transform pair with the conjugate variables $t$ and $n$, with $M_n(u)$ given in Eq.~\eqref{kernel}.  The Green's function problem \eqref{green}
has the following image charge representation:
\begin{equation} \label{image}
\Phi_\mathrm{rf}(\mathbf{r},\mathbf{r}_s)
 =\frac{1}{4\pi \varepsilon_\mathrm{i}}
\int_{r_K}^\infty dx \frac{\rho(x)}{|\mathbf{r}-\mathbf{x}|},
\end{equation}
where the line charge density is $\rho(x) = f(r_K/x)/R$ and $\mathbf{x}=(x/r_s)\mathbf{r}_s$.
\end{theorem}

The image charge representation for the reaction potential Eq.~(\ref{image}) is equivalent to the Kirkwood series representation, Eq. \eqref{kirkwood}. It is advantageous because the line integral can be efficiently discretized by Gauss quadrature.  A few Gauss points can provide approximations with accuracy as high as desired \cite{CDJ:JCP:07}.  Physically, this amounts to approximating line image by a few point images.

\subsection{Point image and line image}
The following limit of the Bessel function can be established (using Eq.~(\ref{kn-aspmptotics}))
\begin{equation}
\lim_{n \rightarrow \infty} \frac{ u \, k_n'(u)}{n \,k_n(u)} = - 1,
\end{equation}
%which has been used to obtain asymptotic image approximations \cite{DC:CCP:07,XDC:JCP:09},
Combining with Eq.~(\ref{kernel}) we obtain the large $n$ limit of the coefficients $M_n(u)$:
\begin{equation}
\gamma \equiv \lim_{n \rightarrow \infty} M_n(u)
 = \frac{\varepsilon - 1}{\varepsilon +1}
= \frac{\varepsilon_\mathrm{i}  - \varepsilon_\mathrm{o}}{\varepsilon_\mathrm{i} + \varepsilon_\mathrm{o}}.
\end{equation}
The coefficients Eq.~(\ref{kernel}) can therefore be decomposed into two parts
\begin{equation}
M_n(u) = \gamma + \delta M_n(u),
\end{equation}
where $\delta M_n(u)$ vanishes as $n$ goes to infinity.

The inverse Mellin transform of a constant $\gamma$ is a delta function $\gamma \delta(t - 1)$, while the inverse Mellin transform of the function $\delta M_n(u)$ is generally a continuous function in the interval $[0,1]$.  The linear charge density therefore can be decomposed into (with $t = r_K/x$):
\begin{eqnarray}
\rho(x ) &=& R^{-1}   \mathcal{M}^{-1}(M_n(u); t)
\nonumber\\
&=&  R^{-1} \gamma \delta(t - 1) +
 R^{-1}  \mathcal{M}^{-1}(\delta M_n(u); t)
\nonumber\\
&=& \frac{\gamma R}{r_s} \delta(x - r_K)
+  R^{-1}  \mathcal{M}^{-1}\left(\delta M_n(u); \frac{r_K}{x}\right).
\end{eqnarray}
The first term corresponds to a point image at the Kelvin point, and the second term corresponds to a continuous line image extending from the Kelvin point to infinity.

The simplest limit is $\varepsilon \rightarrow 0$, where the exterior of the cavity becomes a conductor.  In this limit, we easily see from Eq.~(\ref{kernel}) that $M_n(u) \rightarrow -1 = \gamma$, and   $\delta M_n(u) = M_n(u) - \gamma = 0$.
The image charge distribution reduces to a single point image at the Kelvin point, a well known result.

When the dielectric constants inside and outside the cavity are the same, i.e. $\varepsilon =1$, the point image vanishes, but the line image persists. This is precisely the case of multi-scale reaction field model for electrolytes.

\subsection{Neumann's result revisited}

Let us  review the simple case where there is no screening ions outside the cavity, i.e. $u=\kappa \,R = 0$.
In 1883, Neumann \cite{Neumann:TL:83} found an exact expression for the reaction field, in terms of a point image charge at the Kelvin point and a line image:
\begin{equation}
\Phi_\mathrm{rf} =\frac{\gamma R/r_s}{4\pi\varepsilon_\mathrm{i}|\mathbf{r}-\mathbf{r}_K|}+
\frac{\gamma\varepsilon}{(\varepsilon+1)R}\int_{r_K}^\infty dx \frac{\left(r_K/x\right)^{1/(\varepsilon+1)}}{4\pi\varepsilon_\mathrm{i}|\mathbf{r}-\mathbf{x}|}, \label{neumann}
\end{equation}
where $\gamma=(\varepsilon-1)/(\varepsilon+1)$.    This formula has been re-derived independently by various authors in different fields of applications \cite{Yossel:71,Fin:MB:77,EP:SPSS:85,Pola:QJAM:88,Lindell:RS:92,Norris:IEEP:95}; also see Lindell's review \cite{OL:RSB:03} for the summary of history.
The Mellin transform method was also used by Lindell and collaborators \cite{LES:IEE:92,NL:IEEE:95} for finding image charges of the Poisson equation in layered media.

The Neumann's integral expression can be easily derived from the general result Eq.~(\ref{image}).
%through the finite Mellin transformation method since when
%:
For $u=0$, the harmonic coefficients Eq.~(\ref{kernel}) reduce to,
$$
M_n(0)=\gamma+\frac{\gamma\varepsilon/(\varepsilon+1)}
{n+(\varepsilon+1)^{-1}}.
$$
Its inverse Mellin transform can be exactly calculated using Eq. \eqref{pair}:
\begin{equation}
f(t)=\gamma\delta(t-1)+\frac{\gamma\varepsilon}{\varepsilon+1}t^{1/(\varepsilon+1)}.
\end{equation}
Substituting this back into Eq.~(\ref{image}) we find Neumann's result Eq.~(\ref{neumann}).   %
%Evidently the Dirac delta and the Heaviside functions correspond to the point and line images in the Neumann's expression \eqref{neumann}, respectively.

The application of the Neumann's result in molecular dynamics can be found in two recent papers \cite{LBDXJC:JCP:09,LBSDJC:JCP:11}.  The reciprocal problem of a source charge placed outside of a sphere was also applied in Monte Carlo simulations of colloidal systems \cite{DBL:JCP:11,GXX:JCP:12}.  Historically, what is widely used in computer simulations of biological systems is only single image charge approximations, see works by Friedman \cite{Friedman:MP:75} and Abagyan and Totrov \cite{AT:JMB:94}. These methods are of the first or second order accuracy in the dielectric ratio $\varepsilon = \varepsilon_\mathrm{i}/ \varepsilon_\mathrm{o}$.  They fail to be accurate if the ratio $\varepsilon$ is not small.

\subsection{Limits of large cavity and small cavity}
For $u = \kappa R \neq 0$, the inverse Mellin transform of $M_s(u)$ can not be exactly calculated.  One way to proceed is to use the following asymptotic approximation \cite{XDC:JCP:09},
$$
\frac{uk_n'(u)}{k_n(u)}\approx -\left(n+1+\frac{u^2}{1+u}\right),
$$
which gives the correct leading order asymptotics both for $u\rightarrow0$ and for $u\rightarrow+\infty$.  Substituting it back into Eq.~(\ref{kernel}) leads to
\begin{equation}
M_n(u)\approx \gamma+\frac{\varepsilon-2\varepsilon(1+\tilde{u})/(1+\varepsilon)}{(1+\varepsilon)n+(1+\tilde{u})}, \label{xdc}
\end{equation}
where $\tilde{u}=u^2/(1+u)$.  The inverse Mellin transform of this can be easily found.  This approximation works well for the case of small $u$ (small cavity) and large $u$ (large cavity).

\subsection{Large $n$ asymptotics of $M_n(u)$}
Let us first look at the large $n$ limit of the Bessel function $k_n(u)$.  For sufficiently large $n$, the largest term in the sum Eq.~(\ref{knu-def}) is given by $l = n$.  We can therefore rewrite the summation as
\begin{equation}
k_n(u) = \frac{\pi e^{-u} (2n)!}{(2 u)^{n+1} n! }
\sum_{p = 0}^n \frac{n!(2n-p)!}{(2n)!(n-p)!p!} (2u)^p,
\end{equation}
where we have defined $p = n - l$.  We can expand the function being summed into asymptotic series in terms of $1/n$ and extend the upper limit of summation from $n$ to $\infty$.  The resulting summation then can be calculated order by order in $1/n$.  This can be conveniently done using Wolfram Mathematica.
For example, up to order of $n^{-2}$, we have
\begin{equation}
k_n(u) =
\frac{2^{-n-6} u^{-n-1} \Gamma (2
   n+1)}{\Gamma (n+1)}
  \left(
   32 -\frac{8 u^2}{n}
+ \frac{u^4-4 u^2}{n^2}
      +O\left(n^3\right) \right)
   \end{equation}
It then follows that
\begin{eqnarray}
\frac{u k'_n(u)}{k_n(u)} &=&
-1 - n - \frac{u^2}{4 n^2} - \frac{u^2}{2 n} + O(n^{-3}),
\label{kn-aspmptotics}\\
M_n(u) &=&
\frac{(\varepsilon -1) \varepsilon }{n
   (\varepsilon +1)^2}-\frac{\varepsilon
   \left(u^2 \varepsilon +u^2+\varepsilon
   -1\right)}{n^2 (\varepsilon
   +1)^3}+O\left(n^{-3}\right).
\end{eqnarray}
Generically, therefore,  the leading order term of $M_n(u)$ scales as $1/n$.  In the most interesting case $\varepsilon = 1$, however, this term vanishes and we have
\begin{equation}
M_n(u) = -\frac{u^2}{4
   n^2}+O\left(n^{-3}\right).
\end{equation}
Now consider the limit where the source charge approaches the cavity boundary, we have $r_s \rightarrow R, r_K \rightarrow R$.  The reaction potential acting on the source charge then (see Eq.~(\ref{kirkwood})) becomes
\begin{equation}
\Phi_{\rm rf} (\mathbf{R}, \mathbf{R}) \rightarrow
 \frac{1}{4\pi\varepsilon_\mathrm{i}R}
\sum_{n=0}^\infty M_n(u)  <  \infty.
\end{equation}
Therefore the line image strength must vanish at the Kelvin point.

\subsection{The general case}

\subsubsection{Pad\'e approximations to harmonic coefficients}
In the multi-scale reaction field model, we typically have $\varepsilon \approx 1$ and $u = \kappa R$ of order of unity.  All methods discussed above fail in this case.  To obtain accurate approximation, we approximate the harmonic coefficients $M_n(u)$ by a rational function of $n$ (i.e. {\it Pad\'e approximation}):
\begin{equation}
\tilde{M}(P; n)=
\gamma +    \sum_{j=0}^{P-1}
\alpha_j n^j\left/ \sum_{j=0}^P \beta_j n^j\right.,
\label{rational-app}
\end{equation}
with constants $\alpha_j\in\mathbb{R}$ and $\beta_j>0$ for all $j$.  Clearly Eq. \eqref{xdc} is a special case of Eq.~(\ref{rational-app})  with $P=1$. The second term on the right hand side is {the order $[(P-1)/P]$ Pad\'e approximant} \footnote{Here $P-1$ and $P$ refer to the degree of polynomials in the numerator and in the denominator respectively. } of the function $\delta M_n(u)$, and  the rational polynomial preserves the asymptotics,
\begin{equation}
M_n(u)=\gamma+O(1/n),~~~ \hbox{as}~~ n\rightarrow\infty.
\end{equation}
The coefficients $\{\alpha_j,\beta_j\}$ are solved by the nonlinear least square method. For given $u$ and dielectric ratio $\varepsilon$, the constants in the expansion can be simply determined by a minimization of the total $L_2$ error of the first $N+1$ terms of the harmonic coefficients,
\begin{equation}
\min_{ \{ \alpha_j,\beta_j\} } \sum_{n=0}^N\left[M_n(u)-\tilde{M}(P; n)\right]^2. \label{minim}
\end{equation}
The Newton iteration scheme or other iteration algorithms can be applied to solve this nonlinear optimization problem.

To demonstrate the quality of Pad\'e approximation,  we list in Table \ref{error1} the relative errors for the cases of $P=2$ and 3 with various parameters $u$ and $\varepsilon$:
$$
E^2=\displaystyle \sum_{n=0}^N\left[M_n(u)-\tilde{M}(P; n)\right]^2\left/\sum_{n=0}^N M_n(u)^2\right.,
$$
where $\tilde{M}(P; n)$ is the numerical solution of the nonlinear least square problem \eqref{minim}.  We take $N=50$ since 51 multipoles in the Kirkwood series already provide sufficiently high accuracy. It should be pointed out that the minimization solution depends on the initials and may be not a global minimization.
Nonetheless, the results in Table \ref{error1} clearly show the remarkable precision of the Pad\'e approximation.  For example, the worst case for $P=3$ has a relative error $\sim 0.02\%$. In comparison, the asymptotic method Eq.~(\ref{xdc}) yields much larger error, and completely breaks down for $\varepsilon\rightarrow 1$.

\begin{table}[h]\begin{center}
\caption{Relative errors of Pad\'e approximations to the harmonic coefficients for different parameters $u$ and $\varepsilon$. In the minimization,
$N=50$ is taken. The asymptotic solution Eq.~(\ref{xdc}) \cite{XDC:JCP:09} is also given for comparison.}  \label{error1}
\vspace{2.5mm}
\begin{tabular}{ccccc}\hline

   & $\varepsilon=0.02$   & $\varepsilon=0.1$  & $\varepsilon=0.5$  &    $\varepsilon=1$              \\\hline
  $P=2$            &   & & &                     \\
  $u=0.5$ & 4.10E-6  & 2.03E-5  & 1.10E-4  &  6.01E-4         \\
  $u=2$   & 3.14E-5  & 1.64E-4  & 1.15E-3  &  5.07E-3         \\
  $u=10$  & 6.08E-5  & 3.18E-4  & 2.08E-3  &  5.74E-3         \\    \hline \hline
  $P=3$            &   & & &                     \\
  $u=0.5$ & 3.56E-9  & 2.17E-8  & 4.25E-7  &  1.16E-5         \\
  $u=2$   & 9.55E-7  & 4.92E-6  & 3.31E-5  &  1.41E-4         \\
  $u=10$  & 2.22E-6  & 1.15E-5  & 7.37E-5  &  1.99E-4         \\    \hline  \hline
  Asymptotics Eq.~(\ref{xdc})           &   & & &                     \\
  $u=0.5$ & 1.12E-3  & 5.84E-3  & 4.28E-2  & 0.42           \\
  $u=2$   & 1.67E-3  & 9.34E-3  & 9.05E-2  & 0.58          \\
  $u=10$  & 8.89E-3  & 4.66E-2  & 0.32    &  0.95         \\    \hline
\end{tabular}
\end{center}
\end{table}

\subsubsection{Image expressions through inverse Mellin transforms}

\begin{figure}[htbp!]
\centering\includegraphics[scale=.5]{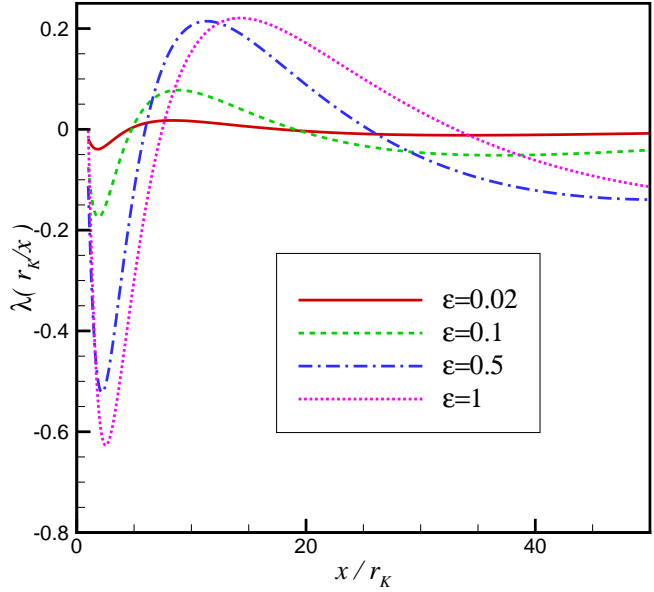}
\caption{ Profiles of line image strengths as a function of $x/r_K$ for parameters $u=2$ and different dielectric ratios $\varepsilon$ calculated with $P=3$.  Note that the linear charge density is always oscillatory in radius.  } \label{lineprofiles}
\end{figure}

After the best-fitting coefficients $\{\alpha_j,\beta_j\}$ are determined, we can further
re-express the Pad\'e approximation Eq.~(\ref{rational-app}) using the {\it partial fraction expansion}:
\begin{equation}
\tilde{M}(P; n) = \gamma
+\sum_{j=1}^P\frac{C_j}{n + Z_j},
\label{partial fraction}
\end{equation}
{where the coefficients $Z_j$ and $C_j$ can be determined by the Heaviside's cover-up method.}
Using Eq. \eqref{pair}, the inverse Mellin transform $f(t)=\mathcal{M}^{-1}[\tilde{M}(P; s); t]$, we easily find the image charge distribution:
\begin{equation}
\rho(x) = \frac{\gamma R}{r_s} \delta(x - r_K)
+ \lambda(r_K/x),
\end{equation}
where
\begin{equation}
\lambda(t) =  \frac{1}{R}
\sum_{j=1}^P C_j t^{ Z_j} \label{lambda}
\end{equation}
with $t=r_K/x$ represents a line image density from the Kelvin image point to the infinity along the radial direction.  Alternatively, the reaction potential is,
\begin{equation}
\Phi_\mathrm{rf}(\mathbf{r},\mathbf{r}_s)
= \frac{\gamma r_K}{4\pi R
\varepsilon_\mathrm{i}|\mathbf{r}-\mathbf{r}_K|}+
\frac{1}{4\pi \varepsilon_\mathrm{i}}\int_{r_K}^\infty dx
\frac{\lambda(r_K/x)}{|\mathbf{r}-\mathbf{x}|}.  \label{lineimage}
\end{equation}
Figure \ref{lineprofiles} illustrates the profiles of line image strengths $\lambda(r_K/x)$ for $u=2$ and for $\varepsilon$ varying from $0.02$ to 1 calculated with $P=3$.  Note that the line image is always oscillatory, which implies that some of the parameters $Z_j$ are complex numbers.  Note also that  both amplitude and period increase with the dielectric ratio $\varepsilon$.

\subsubsection{Discretization of the line image}

Computation of the line integral in Eq.~(\ref{lineimage}) using the continuous linear charge density Eq.~(\ref{lambda}) is still expensive.  This is a quite severe limitation on the computational efficiency, since the line integral has to be computed in every simulation step.
%:
% \xu{in dynamical computer simulations, since this integral is required to compute for each pairwise interaction. A simple approximation may still lead to a $O(N^2)$ complexity with big prefactor.}
%:
Therefore we further discretize the line integral \eqref{lineimage} using numerical quadrature.  This amounts to approximating the linear image using multiple point image charges. {An efficient discretization scheme uses fewer point images, and for large-scale systems the computation of pairwise interactions of these source-image charges can be speeded up with fast multipole-type algorithms \cite{BH:Nature:86,DK:JCP:00,GR:AN:97,YBZ:JCP:04} to achieve approximately linear complexity,}  significantly reducing the computational cost in computer simulations.

We consider the case of $P=3$. Extension of the algorithm to other values of $P$ is straightforward. Since $\tilde{M}(P; n) $ must be real and the line charge is oscillatory, the parameters $\{Z_1,Z_2,Z_3\}$ in Eq.~(\ref{partial fraction}) are generally composed of a positive real number and a complex conjugate pair.   The same clearly also holds for the set of parameters $\{C_1,C_2,C_3\}$.  Let $Z_1,C_1$ be real and
\begin{equation}
Z_{2,3}= p\pm iq, \quad C_{2,3}= a\pm ib.
\end{equation}
The line image strength in Eq. \eqref{lambda} can be expressed as
\begin{equation}
\lambda(t) \, R=C_1\, t^{ Z_1}+2\, t^{ p}
\left[a\, \cos(q\log t)+b\,\sin(q\log t)\right],
\label{lambda-example}
\end{equation}
where $t=r_K/x$.  Note that the second term is oscillatory, in agreement with Fig.~\ref{lineprofiles}.

Approximating an oscillating integral is tricky.  We divide the integral in Eq.~(\ref{lineimage}) into two parts:
 \begin{equation}
 \int_{r_K}^\infty \frac{\lambda(r_K/x)}
{|\mathbf{r}-\mathbf{x}|} dx
= \int_{r_K}^{r_KT} \frac{\lambda(r_K/x)}
{|\mathbf{r}-\mathbf{x}|} dx
 + \int_{r_KT}^\infty\frac{\lambda(r_K/x)}
{|\mathbf{r}-\mathbf{x}|}dx ,
 \label{divide}
 \end{equation}
with $T$ a positive number sufficiently larger than unity.  The first integral in Eq. \eqref{divide} can be transformed into an integral over the interval $v\in[-1,1]$ through the linear variable transformation:
$$x(v)=r_K (2+\epsilon)^\tau/(1+\epsilon-v)^\tau$$
with $\epsilon=2/(T^{1/\tau}-1)$ and $\tau$ a positive constant.  {Due to the finite interval being integrated, the integrand is only weakly oscillatory.}  Further defining a function $Q(v)$ via
\begin{equation}
Q(v) =
{\lambda(r_K/x(v))}
\frac{\tau x(v)}{1+\epsilon-v},
\label{weight-function}
\end{equation}
we can easily show that
\begin{eqnarray}
\int_{r_k}^{r_KT}\frac{\lambda(r_K/x)}
{|\mathbf{r}-\mathbf{x}|}dx
= \int_{-1}^1  \,
 \frac{Q(v)}{|\mathbf{r}-\mathbf{x}(v)|} dv.
\end{eqnarray}
The resulting integral over $v$ can be then integrated using the classical $M$-point Gauss-Legendre quadrature,
which discretizes the line image into several point image charges
\begin{equation}
\int_{r_k}^{r_KT}\frac{\lambda(r_K/x)}
{|\mathbf{r}-\mathbf{x}|}dx \approx
\displaystyle\sum_{m=1}^M
 \frac{q_m}{|\mathbf{r}-\mathbf{r}_m|},
\end{equation}
where
\begin{equation}
\mathbf{r}_m=\mathbf{x}(s_m), \quad
q_m = \omega_m \, Q(s_m)
\nonumber
\end{equation}
%\lambda[r_K/x(s_m)]\frac{\omega_m\tau x(s_m)} {(1+\epsilon-s_m)},$$
and $\{s_m,\omega_m, m=1,\cdots,M\}$ are the Gauss quadrature points and weights.

For the second integral in the RHS of Eq. \eqref{divide}, {the Gauss quadrature is less efficient due to the high oscillatory integrand for small $t$}. Fortunately, the variable $x$ is much larger than $r$, hence the integrand can be expanded in terms of the ratio $r/x$:
$$\frac{1}{|\mathbf{r}-\mathbf{x}|}=\sum_{n=0}^\infty\frac{r^n}{x^{n+1}}P_n(\cos\theta).$$
Therefore
\begin{equation}
\int_{r_KT}^\infty \frac{\lambda(r_K/x)}
{|\mathbf{r}-\mathbf{x}|}dx
= \frac{1}{R} \sum_{n = 0}^{\infty} I_n \, r^n \, P_n(\cos\theta),
\end{equation}
where
$$
I_n=\int_{r_KT}^\infty \frac{\lambda(r_K/x)}{x^{n+1}} dx.
$$
The first few terms can be explicitly worked out using Eq.~\eqref{lambda-example}:
\begin{eqnarray}
&&I_0= \frac{C_1T^{-Z_1}}{Z_1}+2
\frac{(ap-bq)\cos(q\log T)-(aq+bp)\sin(q\log T)}{T^{p}(p^2+q^2)},  \nonumber \\
&&I_1=\frac{1}{r_K}
\left\{ \frac{C_1T^{-Z_1-1}}{Z_1+1}+2
\frac{[a(p+1)-bq]\cos(q\log T)-[aq+b(p+1)]
\sin(q\log T)}{T^{p+1}[(p+1)^2+q^2]}]\right\}
,\nonumber\\
&&I_2=\frac{1}{r_K^2}
\left\{ \frac{C_1T^{-Z_1-2}}{Z_1+2}+2
\frac{[a(p+2)-bq]\cos(q\log T)-[aq+b(p+2)]
\sin(q\log T)}{T^{p+2}[(p+2)^2+q^2]}\right\}.\nonumber \\
\end{eqnarray}
%For sufficiently large $T$, the zeroth order correction $I_0$ already gives an approximation with high accuracy.

In summary, the reaction potential Eq.~(\ref{lineimage}) is approximated by $M+1$ image point charges plus a few correction terms,
\begin{equation}
\Phi_\mathrm{rf}=\frac{1}{4\pi\varepsilon_\mathrm{i}}
\sum_{m=0}^M\frac{q_m}{|\mathbf{r}-\mathbf{r}_m|}
+\frac{1}{4\pi \varepsilon_\mathrm{i}R}
\sum_{l=0}^L I_l \, r^l \, P_l(\cos\theta), \label{asum}
\end{equation}
where the term of $m=0$ represents the Kelvin image charge, with
$q_0 = \gamma r_K/R$ and ${\mathbf r}_0 = {\mathbf r}_K$.  As pointed out previously, the Kelvin image vanishes when $\varepsilon=1$.

{\noindent {\bf Remark.} The error of approximating Eq. \eqref{divide} comes from two sources. One is from the Gauss quadrature to the finite integral, and the other is from the truncation of multipoles for the infinite integral. Both errors depend on the cutoff parameter $T$. In approximating infinite integral, the leading term in the error of the truncation is $O\left[\left(r/r_K\right)^{L+1}\cdot(1/T)^{L+2}\right]$ where $r/r_K<1$. The Gauss quadrature has a fast convergence, hence we expect a small $T$ such that a few points leading to high accuracy. We find $T=4\sim6$ provides a good balance between two approximations in an accuracy of 3 digits.
}

\section{Numerical results}
In this section, we compute the reaction potential using our method and quantify the errors.  We also demonstrate the power of multi-scale modeling by Monte Carlo simulating a dilute symmetric electrolyte.

\subsection{Accuracy and efficiency of discretization scheme}
We first test the accuracy of our discretization scheme.  We focus on the most difficult case $\varepsilon=1$ where the line image is strongly oscillatory, and the point image vanishes.  It is also the case appears in the multiscale modeling of electrolytes. We take $u=5$ and $R=1$, and calculate the relative error of the self energy of a unit charge, $\Phi_\mathrm{rf}(\mathbf{r}_s,\mathbf{r}_s)/2$, as a function of the source point radius $r_s,$  by comparing with 201-terms-truncation of the Kirkwood series. {The latter has a truncation error about $(r_s/r_K)^{201}$ (for example, when $r_s=0.95R$ the error is about $10^{-9}$ ) and can be considered as the ``exact" solution.} We set $\tau=5$ and $T=4$.  We find that the accuracy of numerical quadrature only weakly depends on $T$ and $\tau$.

\begin{figure*}[h!]
\centering\includegraphics[scale=.35]{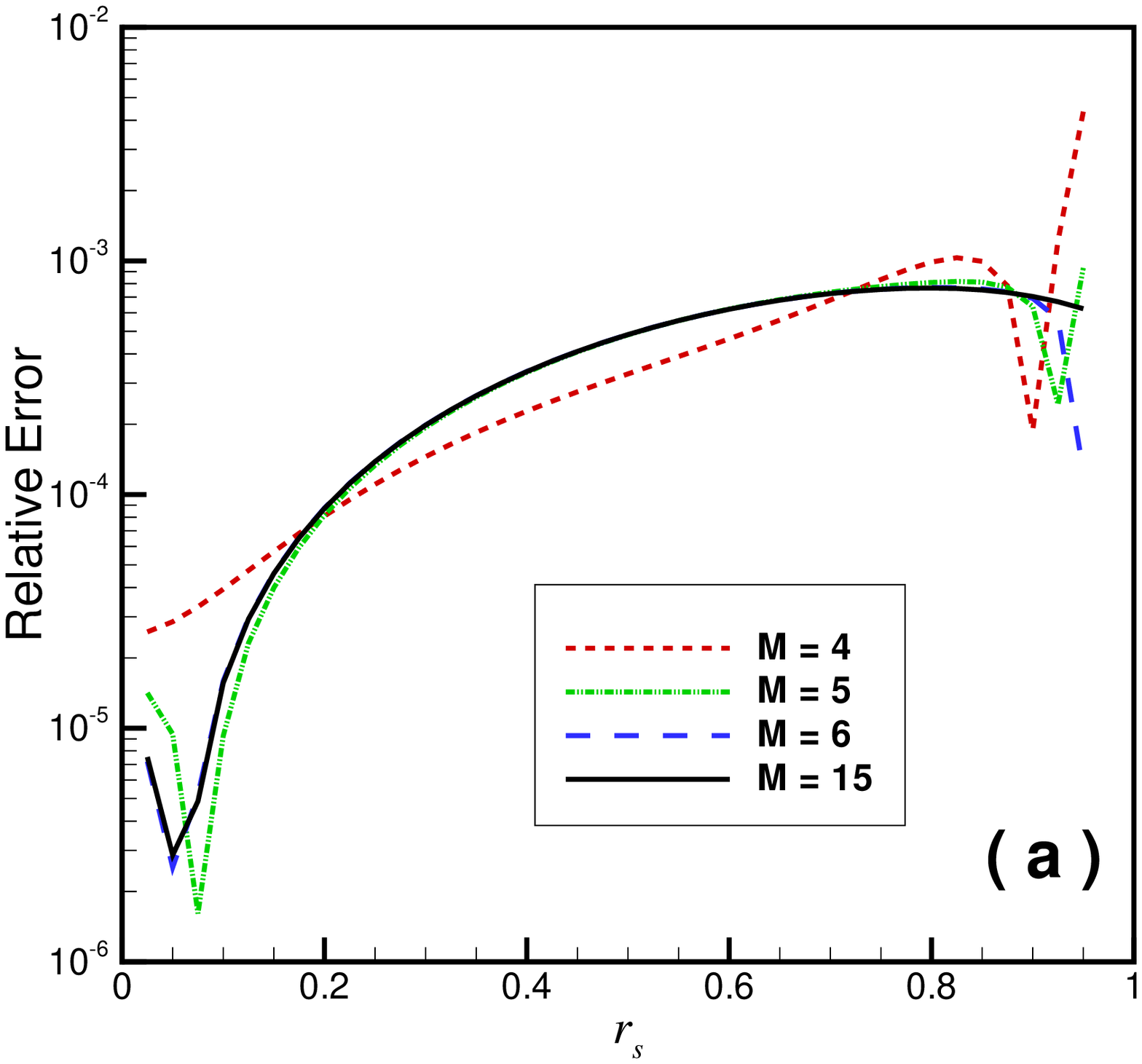}\includegraphics[scale=.35]{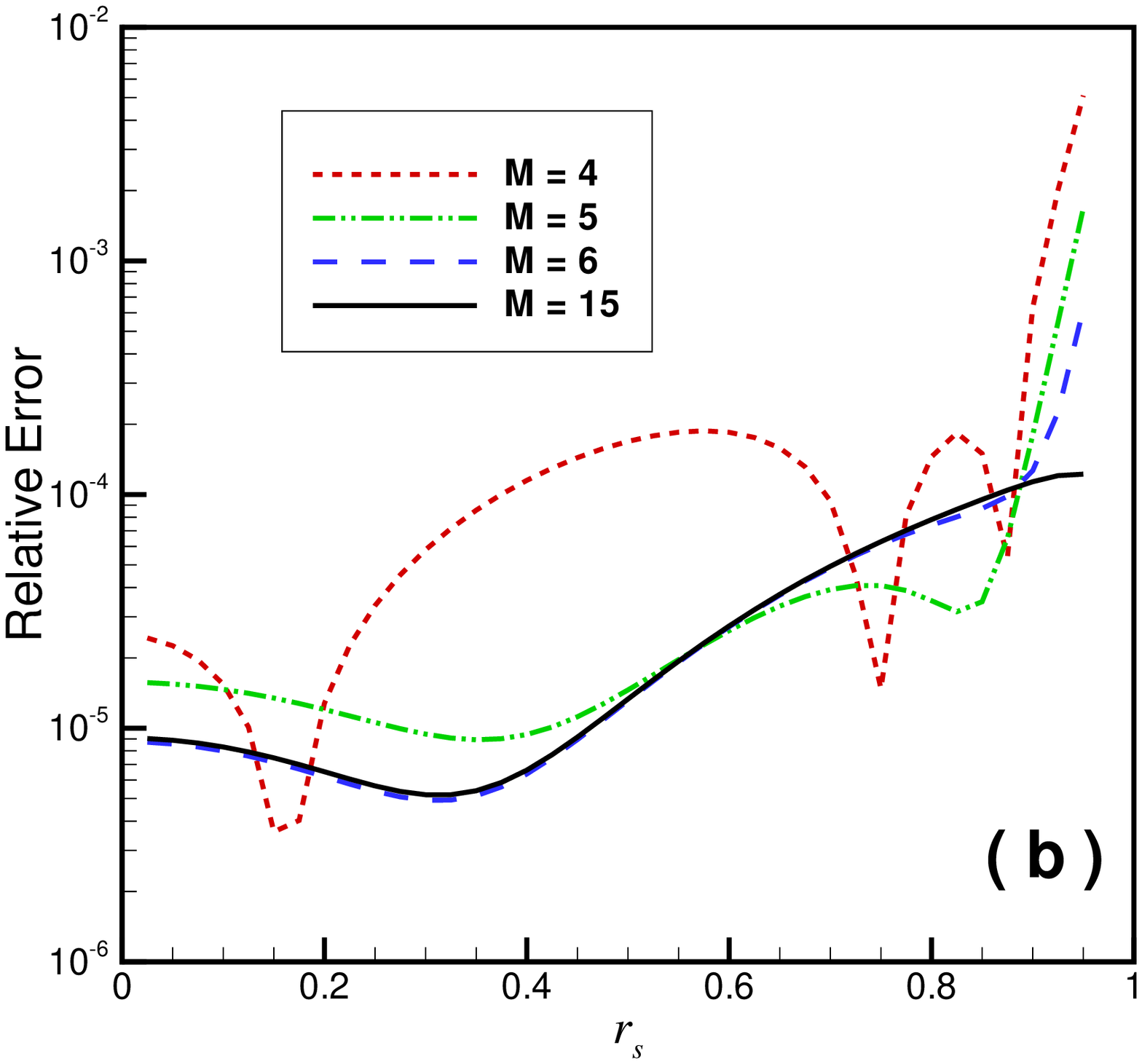}
\caption{ Relative errors of the self energy as a function of source position $r_s$ for different numbers of images ($4, 5, 6, 15$). (a) $L = 0$; (b) $L = 1$. } \label{accuracy}
\end{figure*}

We use $4, 5, 6,$ and $15$ Gauss quadrature points for the integration on the finite interval $[r_K, r_KT]$ and the two different corrections for the integration on interval $[r_KT,\infty)$: $L = 0$ and $1$. Note that $L=0$ means the second term of Eq. \eqref{asum} is a constant correction. The results are shown in Figure \ref{accuracy}. It is seen that the image charge approximation is very accurate even with only 4 image points, with the overall relative errors remaining less than $1\%$.  The accuracy however does decrease when $r_s$ approaches to the boundary.

\begin{figure}[h!]
\centering\includegraphics[scale=.5]{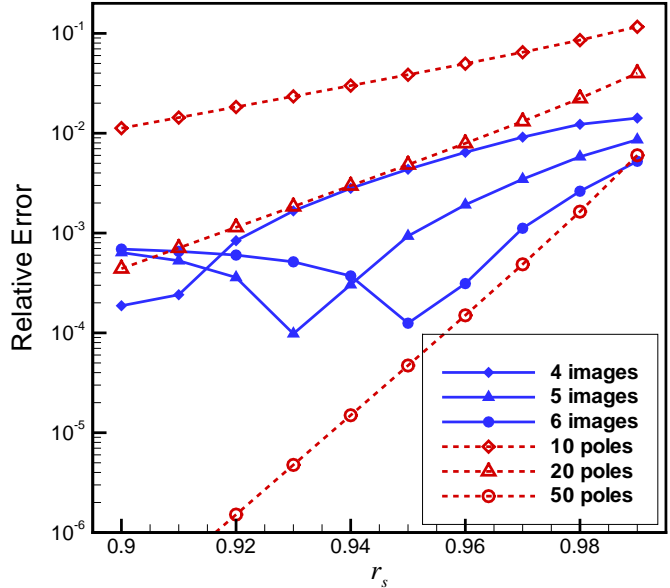}
\caption{ Relative errors of the self energy as a function of source position $r_s\in[0.9,0.99]$ for different numbers of images ($4\sim6$) with constant correction $L=0$ and different numbers of poles with multipole expansion.   } \label{errorcomparison}
\end{figure}

To determine the asymptotics of image charge approximation for the self energy near the boundary, we also compare it with the direct truncation of the Kirkwood series, in the range $r_s/R\in[0.9,0.99]$.   These results, shown in Figure \ref{errorcomparison}, clearly demonstrate that
image charge approximation converges much faster than the Kirkwood series near the boundary.  In particular, image charge approximation with 4 image charges is uniformly better than Kirkwood series with 20 terms.  For example, the error of Kirkwood series with 20 terms is larger than $1\%$ for $r_s/R=0.97$, and increases to $4\%$ for $r_s/R=0.99$, while the error of the image charge approximation remains less than $1\%$.
%These results clearly demonstrate the fast convergence and high accuracy of the approximation with the increase of the image number.

%:
We also compare the CPU timing efficiency of our image charge method and that of the Kirkwood series. %In the aspect of the CPU timing, the image method is very competitive in comparison to the Kirkwood series.
%:
The comparison was performed using the optimized module for the Bessel functions by Matlab, and the algorithm of the Legendre polynomials in Numerical Recipes \cite{PTVF:book:92}. The machine used has double-core 2.3GHz CPU and 8G memory. We use the same system parameters as in the accuracy test and calculate the pairwise energy of $N$ ions randomly distributed in the cavity. In the image method we use $4\sim6$ images and $L=0$ and 1 corrections, respectively.  The Kirkwood series are truncated at 10th, 20th and 50th terms. The results are listed in Table \ref{CPU}.  We see the image method with 4-point images is generally 40-50 times faster than the method of Kirkwood series truncated at 20-th terms.
With these parameters, two methods have similar accuracy.

%the computation is speeded up by a factor of 40 to 50 if using the image method from the comparison between the 4-point image method and the 20-term spherical harmonics, for which the two has similar accuracy.

\begin{table}[h]\begin{center}
\caption{The CPU timing for comparison between image charge method and the truncation of the spherical harmonics series. The time unit is seconds.}  \label{CPU}
\vspace{2.5mm}
\begin{tabular}{c|ccc|ccc|ccc}\hline
\multirow{2}{*}{$N$}    &  \multicolumn{3}{c|}{Images, $L=0$}   &   \multicolumn{3}{c|}{Images, $L=1$}    &    \multicolumn{3}{c}{Spherical harmonics}  \\
     &  4    &   5    &  6    &   4    &   5   &    6    &   10    &   20   &   50       \\ \hline \hline
50  &   0.058    &   0.066  &   0.075   &   0.067   &   0.077  &    0.082   &   1.36    &   2.81   &   8.81       \\
100 &   0.22    &   0.28    &   0.31    &   0.26    &   0.31   &    0.34    &   5.29    &   11.40   &   35.54       \\
200 &   0.86    &   0.97    &   1.11    &   1.01    &   1.15   &    1.26    &   21.28   &   45.30   &   144.8        \\    \hline
\end{tabular}
\end{center}
\end{table}

\subsection{Application in Monte Carlo simulations of electrolytes}
To illustrate the utility of our image charge method, we apply it in a reaction-field Monte Carlo simulation of electrolyte.  We artificially introduce a spherical cavity with radius $R$ in the electrolyte, and model all charges in the cavity using the primitive model and simulate them using Monte Carlo method.  All ions outside the cavity, together with the solvent, are treated implicitly using linearized PB theory, characterized by two parameters: the dielectric constant $\epsilon_\mathrm{o}$ and the inverse Debye length $\kappa$.  For any ion inside the cavity, the effects of the electrolyte outside the cavity is to introduce a reaction potential, which can be calculated using our image charge method.  It is important to note that the  inverse Debye length $\kappa$ characterizing the medium outside the cavity shall be self-consistently determined by the simulation of the ions inside the cavity.

We run canonical Monte Carlo simulations of the primitive model using the standard Metropolis criterion \cite{Metropolis:53,FS:book:02} for particle displacements.  The ions are modeled by hard spheres with diameter $D=3.75 \AA$ and with a point charge of valence $z_i=\pm 1$ at its center.  The ions are mobile in the solvent medium with dielectric permittivity $\varepsilon_\mathrm{i}=80$. The effective Hamiltonian of the system is given by
\begin{equation}
U=\sum_{i=1}^{N} U_{i}^\mathrm{self} + \sum_{i<j}U_{ij}.
\label{hamiltonian}
\end{equation}
The self energy $U_{i}^\mathrm{self} $ is given by
\begin{equation}
\beta U_i^\mathrm{self}=\left\{ \begin{array}{ll}
\frac{1}{2} l_B z_i^2 4\pi\varepsilon_\mathrm{i}
\Phi_\mathrm{rf}(\mathbf{r}_i,\mathbf{r}_i),
\quad\quad& r_i\leq R, \\
\infty,& r_i > R,
\end{array}\right.
\label{self-energy-hamiltonian}
\end{equation}
where $l_B=e^2/(4\pi\varepsilon_0\varepsilon_\mathrm{i}k_BT)$ is the Bjerrum length and $\beta=1/(k_BT)$ is the Boltzmann factor. We take $l_B=7.14\AA$  for the water permittivity at room temperature. The infinite potential is due to the presence of the hard wall at $r=R+D/2.$ The pairwise interaction energy $U_{ij}$ is given by
\begin{equation}
\beta U_{ij} =  l_B z_iz_j \left[
\frac{1}{|\mathbf{r}_i - \mathbf{r}_j| }
+4\pi\varepsilon_\mathrm{i} \Phi_{\mathrm{rf}}(\mathbf{r}_i, \mathbf{r}_j)
\right],
\end{equation}
As is well known, the reaction potential is symmetric under permutation of two variables: $\Phi_{\mathrm{rf}}(\mathbf{r}_i, \mathbf{r}_j) = \Phi_{\mathrm{rf}}(\mathbf{r}_j, \mathbf{r}_i)$.

\begin{figure*}[htpb!]
\centering\includegraphics[scale=0.5]{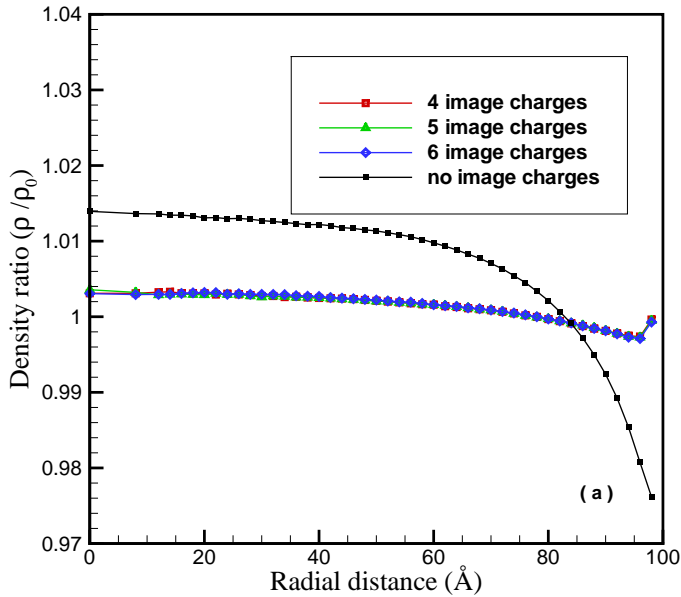}  
\caption{Comparison of radial density ratios of anions in the spherical cavity for approximating the reaction field with different numbers of images, and without reaction field. Cavity radius is $R=100 \AA$; the salt concentration is $\rho_0=8mM$, corresponding to a Debye length $\kappa^{-1} = 34\AA$.} \label{rdf}
\end{figure*}

We first calculate the density distribution by taking a system with $R=100 \AA$ and the salt concentration is $\rho_0=8mM$ which corresponds to a Debye length $\kappa^{-1} = 34\AA$. We calculate both the cases with and without the reaction field.  In the former case, we use different numbers of image charges varying from 4 to 6, with the parameters for numerical quadratures the same as those in the upper panel of Figure \ref{accuracy}.  For each setting, we run $6\times10^8$ MC cycles for each particle to obtain samples for the statistics of particle number in each spherical shell with thickness $2\AA$.

In Figure \ref{rdf} we show the radial distribution function of anions, i.e., the normalized density, $\rho/\rho_0$.  The distribution of cations is similar.   Evidently, without accounting for the reaction field, the density is higher near the center of cavity and lower near the cavity boundary.  The difference in the density is up to 3 percent.   When the reaction field is taken into account, the particle density shows variation less than $0.5\%$ in most regions inside the cavity.  There is substantial deviation of density near the cavity boundary.  This is due to the presence of the artificial hard wall on the boundary, see Eq.~(\ref{self-energy-hamiltonian}).  This problem can be cured by adding to the self energy a short ranged correction, or by introducing a buffer zone of a certain thichness ($\sim D$).  The latter approach has been discussed in literature \cite{AL:JCP:93,WH:JPC:95,LBDXJC:JCP:09}.   We shall present statistical mechanical discussion of the former approach in a separate publication.

\begin{figure}[htpb!]
\centering\includegraphics[scale=0.35 ]{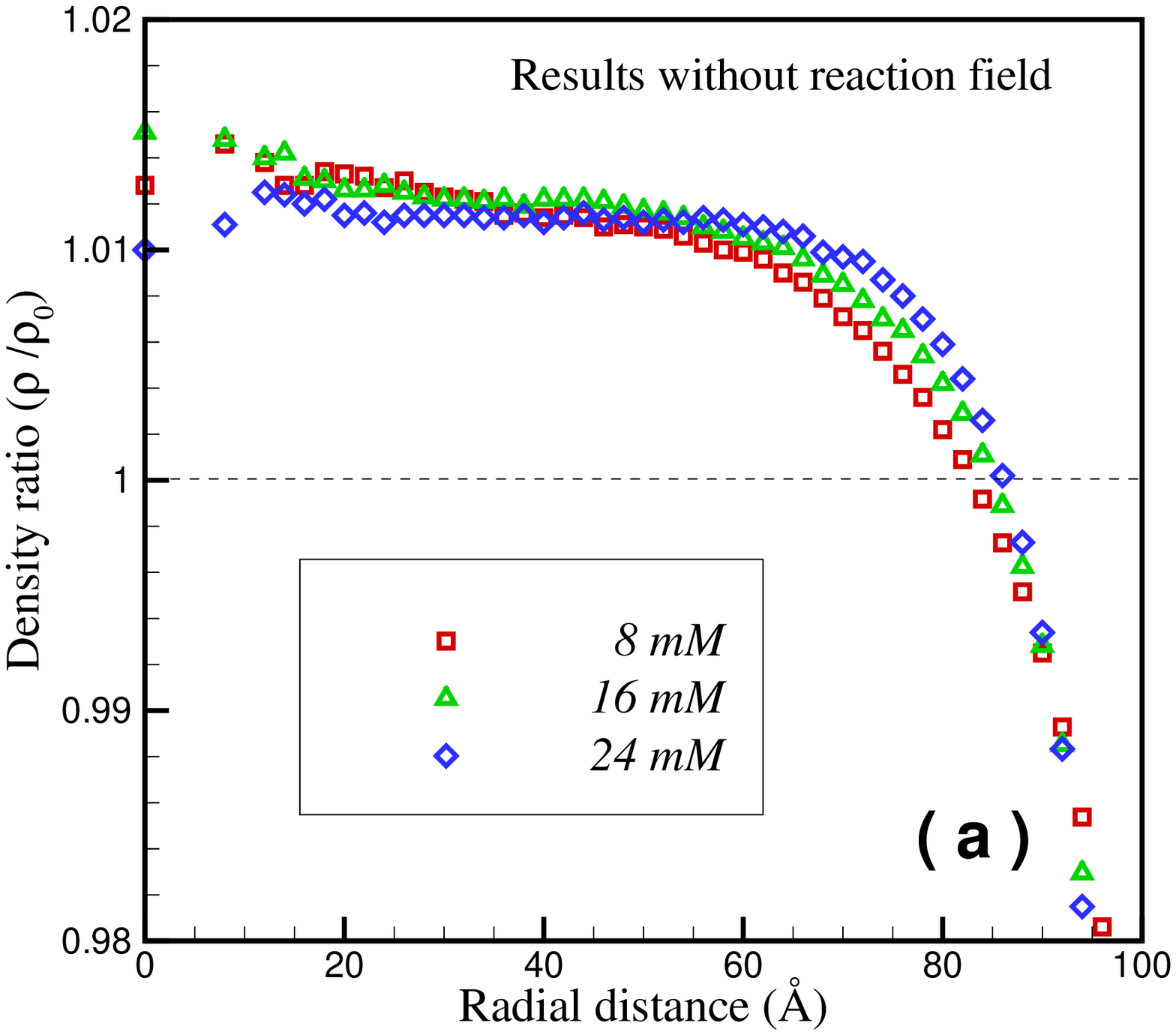}\includegraphics[scale=0.35 ]{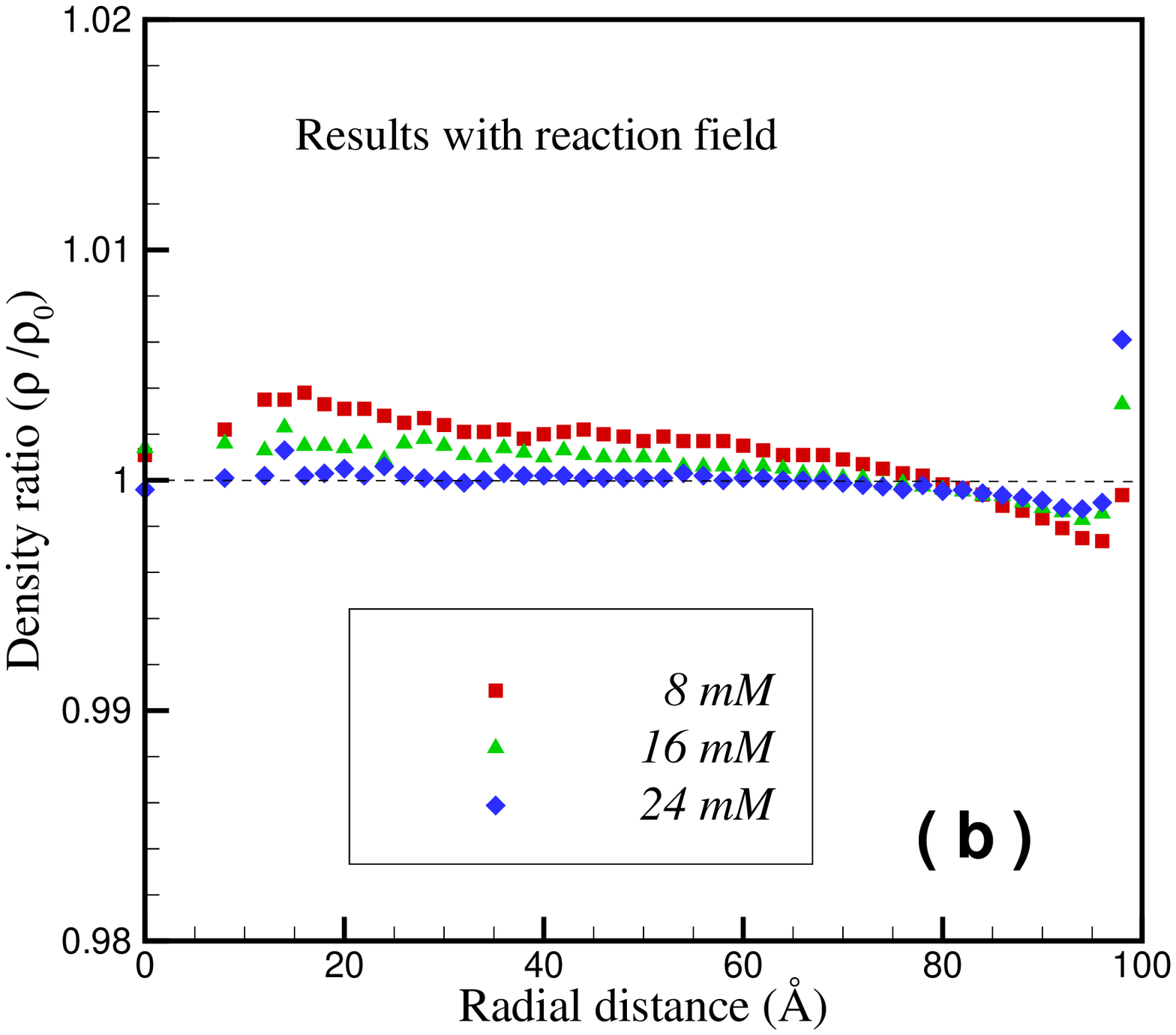}
\caption{ Radial density ratios for three different ion densities, $8mM,16mM, 24mM$ respectively.  The corresponding Debye lengths are $34\AA, 24\AA$ and $20\AA$ respectively.  Top panel: without reaction field;  Bottom panel:  with reaction field.  For computation of the reaction potential, $M = 5$ and $L =0$ are adopted.} \label{rdf2}
\end{figure}

Finally, in Figure \ref{rdf2}, we plot the results for three different bulk ionic concentrations, where $\rho_0= 8, 16$ and $24 mM$ respectively.  These results again show the necessity of treating the reaction potential, as well as the accuracy and efficiency of our image charge methods.  It  is also interesting to note that as the Debye length decreases the density becomes flatter, suggesting that the reaction-field Monte Carlo models becomes more precise.  For the case of salt concentration $24mM$, for example, the Debye length is approximately half of the cavity radius, while the density variation becomes less than $0.1\%$ (excluding the thin region affected by the hard wall artifacts).

\section{Conclusions}

In summary, we have developed an image method for charges inside a spherical cavity that is immersed in an ionic solution.  Our method is useful for multiscale reaction field models of electrolytes and other more complicated charged systems. We derive an analytic expression for the reaction potential in terms of one dimensional image charge distribution, and discuss a highly accurate and efficient algorithm for discretizing the image line charge.  We also apply our method to a reaction field Monte Carlo simulations of $1:1$ electrolytes.  and demonstrate the accuracy and efficiency of the new algorithm.

Simulation of charged systems is computationally expensive, therefore is always limited to small system size.  In a physical system of such size, the total charge may fluctuate
away from zero to a noticeable extent.  These fluctuations can not be taken into account  in canonical ensemble simulations.  In another word, grand canonical ensemble must be used to capture the charge fluctuations of small systems.  Ewald-summation method, which is so far the most popular simulation methods for charged systems,
are based on periodic boundary conditions, and are difficult to be incorporated with grand canonical ensemble.  By strong contrast, reaction-field type of models, besides being more intuitive, can be easily adapted to a grand canonical Monte Carlo simulation.   This shall be the topic of a separate publication.

\section*{Acknowledgments}
The authors acknowledge the financial support from the Natural Science Foundation of China (Grant Numbers: 11101276, 11174196, and 91130012) and Chinese Ministry of Education (NCET-09-0556). Z. X. acknowledges the financial support from the Alexander von Humboldt foundation for a research stay at the Institute for Computational Physics, University of Stuttgart. The authors thank Professors Wei Cai and Chunjing Xie for helpful discussion.

%%%% Bibliography  %%%%%%%%%%
%%{unsrt}%{SIAM}% {abbrv} %{nature} %{plain} %{abbrv} %
%\bibliographystyle{SIAM} %{elsart-num}  %{elsart-num-sort}
%\bibliography{D:/xuzhenli/CurrentProjects/biobib}
%\bibliography{biobib}

\end{document}